\documentclass[conference]{IEEEtran}

\usepackage{pdfpages}
\usepackage{graphicx}
\usepackage{epsf} 
\usepackage{subfig}
\usepackage{algorithm}
\usepackage{algorithmic}
\usepackage{fancyhdr}
\usepackage{listings}
\usepackage{caption}
\usepackage{adjustbox}
\usepackage{graphicx}
\usepackage{pgfplots}
\usepackage[T1]{fontenc}

\ifCLASSOPTIONcompsoc
  \usepackage[nocompress]{cite}
\else
  \usepackage{cite}
\fi

\IEEEoverridecommandlockouts

\begin{document}

\title{Blockchain-based Rental Documentation Management with Audit Support}

\author{
\IEEEauthorblockN{Jo\~{a}o F. Santos$^{1,2}$, Miguel P. Correia$^1$, Tiago R. Dias$^2$}
\IEEEauthorblockA{$^1$Instituto Superior T\'ecnico, Universidade de Lisboa - Portugal\\
$^2$Unlockit.io - Portugal
\\
joao.filipe.santos@tecnico.ulisboa.pt, tiago.dias@unlockit.io, miguel.p.correia@tecnico.ulisboa.pt}
}

\maketitle

\thispagestyle{plain}
\pagestyle{plain}

\begin{abstract}
Document management in the rental market is a critical process to ensure the accuracy of financial transactions and regulatory compliance in the sector. In Portugal, the challenges include the complexity of legislation, particularly GDPR non-compliance, lack of transparency, and bureaucratic process inefficiency. With this in mind, a solution based on Hyperledger Fabric, a blockchain platform, is presented for the implementation of a document management system for the rental process. This system oversees the rental process, which consists of three phases: the application for a property by the prospective tenant through the upload of necessary documents, acceptance/rejection by the landlord of various received applications, and the creation of a report by the system, which only the auditor can request and view. The system smart contract records metadata associated with the documents (hash, owner) and coordinates requests for file access by landlords to prospective tenants. Thus, the system is responsible for creating immutable and traceable records of the entire process. The underlying platform serves as the foundation for conducting future audits. After the landlord verifies the files and accepts the rental proposal, any authorised auditor can request a report for a property by accessing the records through the final report, which includes all events that occurred during the process. 
\end{abstract}

\section{Introduction}

\label{chapter:introduction}

Blockchain technology provides decentralisation to sectors where the single point of failure is the norm \cite{peck2017blockchains,dinh2018untangling}. Developers and entrepreneurs continue to add new use cases for this new technology. The so-called cryptocurrencies, like Bitcoin \cite{narayanan2016bitcoin}, are an excellent illustration of this, since their primary purpose is to provide an alternative payment system that is decentralised but otherwise functions similarly to conventional currencies, creating a more transparent and efficient way to conduct financial transactions. However, centralised organisations such as banks can also look at blockchain technology as an innovation with the ability to improve efficiency and transparency. There are still many use cases where blockchain can enter and improve the field, such as Supply Chain \& Logistics, Finance, and Property \& Real Estate, given that this is a new technology undergoing improvement and development. 

Today, the real estate market is inefficient in responding to consumer needs. To obtain or rent a new house, many layers of bureaucracy must be completed to ensure that everything complies with legal requirements; as a result, these procedures are time-consuming. The processes that slow down this procedure are risk verification, regulatory compliance, and fraud investigation, one of many reasons for real estate sluggishness.

Real estate audits and due diligence processes require manual validation of all documentation. An audit is necessary to investigate information about properties and their owners, whose objective is to form an independent opinion on financial and legal statements. During the process, all financial, accounting, and tax aspects of the property must be analysed. This process seeks to catch any possible fraud that the buyer, seller, or renter may be committing, for instance, debts that somehow may be linked to the property itself or even fake property reviews and forgers, who are armed with false documentation, to impersonate owners, sellers, or even attorneys. High intermediary or brokerage fees, the acquisition and verification of pertinent information from legal sources, fluctuating transaction prices, opacity of property rights, and tax fluctuation are other relevant issues within this entire procedure in real estate \cite{importance_audit}. Due to its complexity and difficulty in regulation, it may be difficult to create innovative solutions to solve this real estate audit problem. 

Blockchain shows a promising future for the enhancement and development of new tools in this area. However, it is imperative to recognise that these technologies are not bulletproof. As a result, in addition to the benefits mentioned earlier, the space may suffer from any drawbacks these technologies may have. In the future, auditors may use those techniques to make the audit process more efficient and provide a better and more sophisticated service to anyone seeking their services.


\section{Background}
\label{chapter:background}

A blockchain is a distributed, durable, and append-only ledger that contains records organised into blocks. Blocks store valid transactions as a record book page. It is a distributed ledger technology (DLT) in which a central authority does not maintain the ledger. This technology can be used to ensure safe transactions, reduce compliance costs, and simplify data processing. When a block is filled with transactions (block $t+1$), it is closed and linked to the previously served block (block $t$), forming a chain of data, the blockchain. Following that newly added block, all additional data is collected into a new block, which is then added to the chain once it is complete. The transaction validation process ensures that transactions and blocks of the blockchain are verified \cite{narayanan2016bitcoin}.

\subsubsection{Smart Contracts}
\label{section:smart-contracts}

Smart contracts transform the way agreements are made and enforced \cite{ethereum,zheng2020overview}. These self-executing digital agreements operate autonomously within blockchain networks, automating and verifying contract terms. They are coded to execute actions based on predefined conditions, whether simple criteria or complex events. Crucially, they interact with the blockchain data and modify it as needed.

These contracts are integral to decentralised blockchain networks and leverage low-level programming languages such as Ethereum bytecode. They find applications in various domains, offering transparency, automation, and security. By eliminating intermediaries and reducing manual intervention, they improve efficiency and accuracy, making them valuable for supply chain management, finance, and real estate. Strong encryption and cryptographic techniques enhance the security of smart contracts.

\subsubsection{Permissionless Blockchains}

In permissionless blockchains, there are no restrictions to join the network \cite{peck2017blockchains}. Anyone can participate in the consensus algorithm and validate the data. A user generates a personal address on a permissionless blockchain and interacts with the network by sending transactions to other users or assisting the network in validating transactions. If a user helps with block validation, a reward is earned for validating the new incoming blocks, so this type of blockchain receives more support from the community. In simple terms, it is an entirely decentralised blockchain platform between unknown parties.

\subsubsection{Permissioned Blockchains}

A permissioned blockchain restricts access to authorised users and is often chosen for enhanced security. Blockchain administrators manage user authorisations, ensuring that only authorised individuals can interact with them. This approach is popular among organisations that prioritise data security and anonymity. Permissioned blockchains are widely adopted, especially in corporate settings.

Various permissioned blockchain frameworks, such as Canton \cite{canton}, Hyperledger Fabric \cite{hyperledger_fabric}, and R3 Corda \cite{r3_corda}, support programmable transactions, allowing entities managing the blockchain to define business rules and logic. 

\subsubsection{Hyperledger Fabric}

Hyperledger Fabric, part of the Hyperledger project under the Linux Foundation, is a permissioned distributed ledger platform. It emphasises modular architecture and adaptability, supporting various consensus algorithms. Fabric unique feature is its use of channels for segregated communication paths, ensuring data privacy. Customisable endorsement policies streamline transaction agreement, focussing on scalability. Identity management and access control are maintained through the Membership Service Provider (MSP) framework, ensuring accountability and transparency. Transactions in Fabric undergo a lifecycle, including proposal, endorsement, block distribution, consensus, and ledger update. 

\subsubsection{R3Corda}

R3 Corda is an enterprise-grade distributed ledger platform that focusses on privacy, scalability, and interoperability. It emphasises shared ledgers, where only the parties involved access the transaction data for confidentiality. Fine-grained permissions determine data access. Corda Flow Framework allows direct communication and negotiation, similar to real-world agreements. Smart contracts, called states and contracts, govern shared facts and transaction rules. Pluggable consensus models enable the choice of the most suitable algorithm for security and performance. 

\subsubsection{Canton}

Canton is an innovative blockchain technology that prioritises efficiency, scalability, and practicality. It employs the Proof-of-Stakeholder (PoSH) consensus mechanism, considering participants' roles for influence, ensuring decentralisation and efficiency. Canton's architecture focusses on scalability using infrastructure nodes that efficiently process and verify transactions without accessing data content directly. Secure communication within defined trust zones optimises network efficiency and supports high-throughput use cases. By combining PoSH with an efficiency-oriented design, Canton offers an enterprise-ready platform for blockchain implementation. More details can be found in \cite{canton}.


\section{Audit Process in Real Estate}
\label{chapter:related_work}

An audit is a meticulous process that verifies how well an organisation complies with various demanding requirements, which can be regional, national, or international in nature \cite{importance_audit}. These requirements can vary significantly based on the organisation's location, like tax obligations. Real estate audits are comprehensive, examining financial records, transaction procedures, and document quality. Financial scrutiny involves a detailed review of all money flows within a fiscal year, ensuring the accuracy of financial records. In addition, real estate audits focus on ensuring that transactions comply with local real estate laws. This includes property transactions, leases, contracts, and agreements. The quality of recordkeeping is crucial, as it demonstrates an organisation's commitment to regulatory standards. 

\subsection{Audit Categories}

Audits are divided into three main categories: internal, external, and governmental audits. In internal audits, company employees often conduct internal audits. However, the business can also choose to contract out this service. In external audits, for the audit process to be impartial, external auditors, unlike internal auditors, must be able to operate on their own and provide an unqualified opinion. The last is that government audits are performed to verify that financial statements have been made appropriately and that a company's taxable income has not been distorted \cite{audit_internal_external}. For the aforementioned reason, audits are essential for business continuity \cite{audit_objectives}:

\begin{itemize}
   
   \item \textit{Increase operational efficiency:} Find control suggestions to increase the efficacy and efficiency of processes by regularly assessing and monitoring them.

   \item \textit{Evaluate risks and protects assets:} Assist in keeping track of any environmental alterations documented, as well as ensuring that any risks discovered are mitigated.

   \item \textit{Assess organisational controls:} Enhances the organisation's control environment by analysing effectiveness and efficiency.

   \item \textit{Ensure legal compliance:} Applicable laws and regulations are followed by conducting internal audits on a regular basis.
   
\end{itemize}

Even with a promising digital transformation in the audit process, many obstacles still exist. Auditors need to acquire the skills to undergo this digital transformation and are not ready to approach a more automated audit workflow, creating a significant obstacle in this regime change.

In relation to this topic, \textit{SmartAudit} is a company that performs smart audits at the highest level. Tasks such as lead scheduling, financial statement preparation, and report writing are automated, and the progress of that audit is seen in real time \cite{smart_audit_company}. When new clients are accepted, they must upload their data to their cloud-based infrastructure. Afterward, a new plan is set to audit all these available files, complying with international standards. In the end, a report is generated.

\subsection{Blockchain-based Smart Audit}

A more reliable and effective environment for auditability is produced by combining blockchain with the above smart audit techniques. Due to the sufficiency, relevance, and dependability requirements for audit evidence, blockchain technology is suitable for use in conjunction with intelligent auditing approaches. The integrity of the data provided by blockchain increases the trustworthiness of the audit evidence. The information flow steps are the following:

\begin{enumerate}

    \item \textit{Data Production and Control} Data is collected using smart sensors, IoT, and other technologies. To find anomalies and useful information, a number of tests and analytics are performed using intelligent audit modules.

    \item \textit{Data Storage} The data is maintained in a selected blockchain, which guarantees the integrity and reliability of data.

    \item \textit{Smart Contract Data Manipulation} Smart contracts enforce, without human intervention, the proper operation of intelligent audit modules.

    \item \textit{Data Auditing} The data stored in the blockchain will then be used to perform an audit of those data with the help of tools such as intelligent process automation, natural language processing, and machine learning.

\end{enumerate}

This interaction offers a long-needed solution to many current problems. It was required to look for different approaches in other industries, adapt those ideas in Section \ref{chapter:architecture}, and develop a more sophisticated solution due to the absence of information and solutions with these technologies. The following section explores the state-of-the-art of the most recent techniques with the information provided by blockchain technology.

\subsection{Related Work}

The adoption of blockchain technology has been increasing in multiple sectors, including real estate. In this discussion, researchers propose specific technical solutions that demonstrate the tangible benefits of blockchain in the real estate industry.

Several research efforts have explored the application of blockchain technology in various industries, including real estate. Here are some key findings:

Kang et al. \cite{kang2022blockchain} proposed a solution using blockchain, a peer-to-peer network, and the Interplanetary File System (IPFS) to improve the storage and sharing of knowledge files. This approach improves decentralisation, scalability, and data consistency. It combines blockchain, P2P networks, and IPFS to create a secure and efficient system for storing and sharing knowledge files.

Wouda et al. \cite{wouda2019blockchain} discussed the potential of blockchain to streamline commercial real estate asset transactions, particularly in the Netherlands. They highlighted the challenges faced in these transactions, such as high costs and lack of transparency, and proposed the use of blockchain to address these issues. The aim is to create a transparent and efficient infrastructure for real estate transactions.

Bharimalla et al. \cite{blockchain_nlp_india} presented a solution for an Electronic Health Record System using blockchain, natural language processing (NLP), and machine learning. They implemented a permissioned blockchain, Hyperledger Fabric, to manage access to electronic health records. NLP and OCR technology was used to digitise paper medical records, which were then standardised and stored on the blockchain.

Belchior et al. \cite{belchior2020towards} proposed an auditing solution for a critical Portuguese government application, JusticeChain. External oracles provided audit logs to JusticeChain, which were processed and recorded on a permissioned blockchain, Hyperledger Fabric. Auditors could access these logs only with the consensus of most auditors, ensuring the integrity of the audit process.

\section{Blockchain-Based Documentation Management Architecture}
\label{chapter:architecture}

The Blockchain-Based Documentation Management (BDM) Architecture is a system designed to enable users to efficiently manage their real estate document sharing and transactions securely. Within this architecture, users exercise control over their document management through a set of defined processes, leveraging the immutable characteristic of a blockchain for enhanced data security and transparency. The BDM Architecture incorporates two fundamental approaches to streamline document management: proactive consent and consent on request. In the proactive consent approach, users have the ability to proactively grant or withhold consent to share their documents with relevant parties. This consent is encapsulated in permissions, which are subsequently recorded on the blockchain.

\subsection{Solution Overview}
\label{section:The Blockchain-based Platform}

\begin{figure}[h]
\centering
\includegraphics[width=0.5\textwidth]{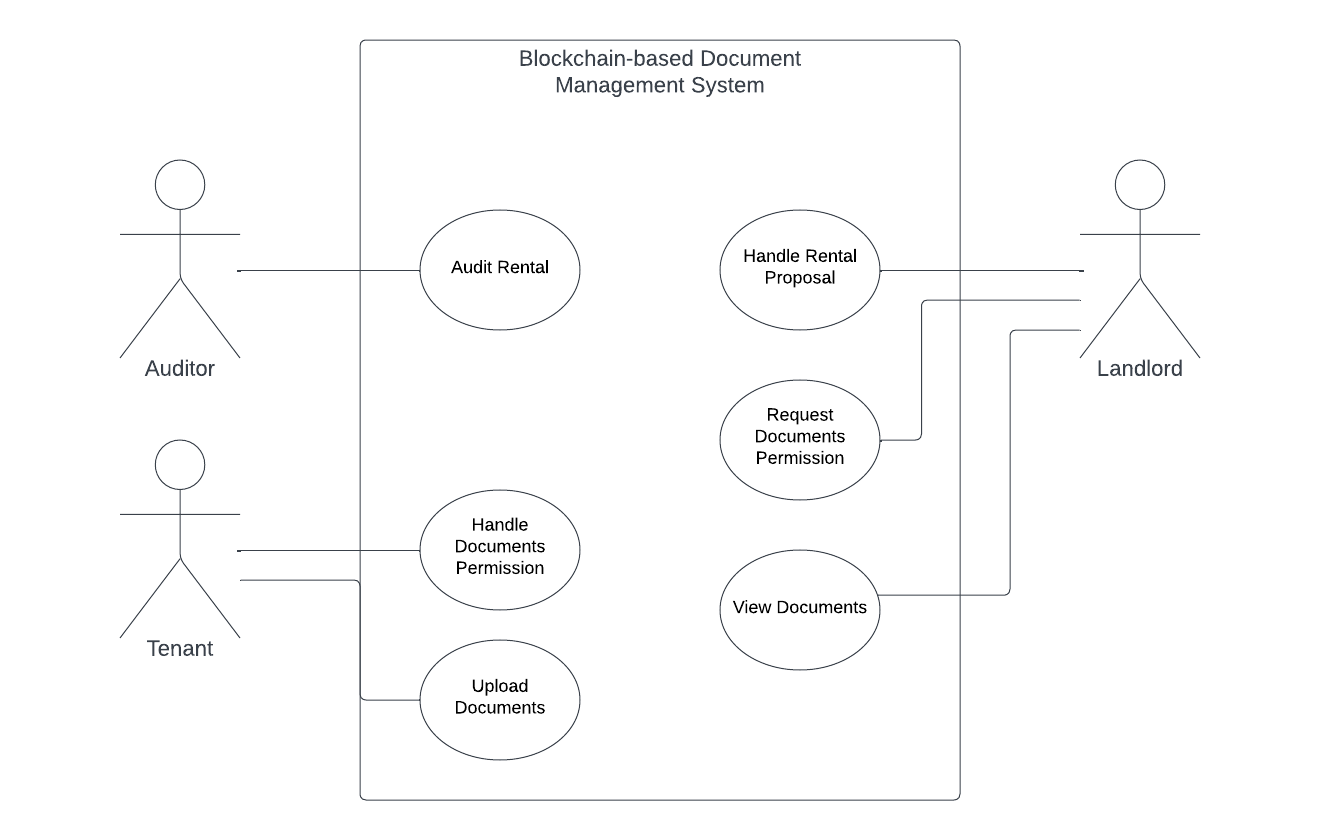}
\caption{BDM Use Case Diagram}
\label{use_cases}
\end{figure}

This application involves three primary user roles: the auditor, the tenant, and the landlord, each with their own distinct actions and responsibilities. Figure \ref{use_cases} presents each one of the actions for each role of the participant.

\captionsetup[figure]
{font=footnotesize,labelfont=small,labelfont={bf}}

\begin{itemize}
    \item \textit{Tenant:} The tenant can upload documents to submit a rental application or can handle the permissions to access the uploaded documents. Those documents and metadata generated from the documents are stored securely for future audit purposes.

    \item \textit{Landlord:} The landlord is an individual or entity who owns and typically manages a property, such as a house or apartment, and rents or leases it to tenants in exchange for a periodic payment. He can handle a rental proposal, either by declining or accepting it. He can also request to see the original uploaded documents. This permission must be granted by the owner of the document.
    
    \item \textit{Auditor:} The auditor can audit any house he chooses and has access to. The BDM will generate a report with each transaction that happened in that house for the auditor to analyse.
    
\end{itemize}

\subsection{System Architecture}
\label{section:BBDM Architecture}

\begin{figure}[h]
\centering
\includegraphics[width=0.5\textwidth]{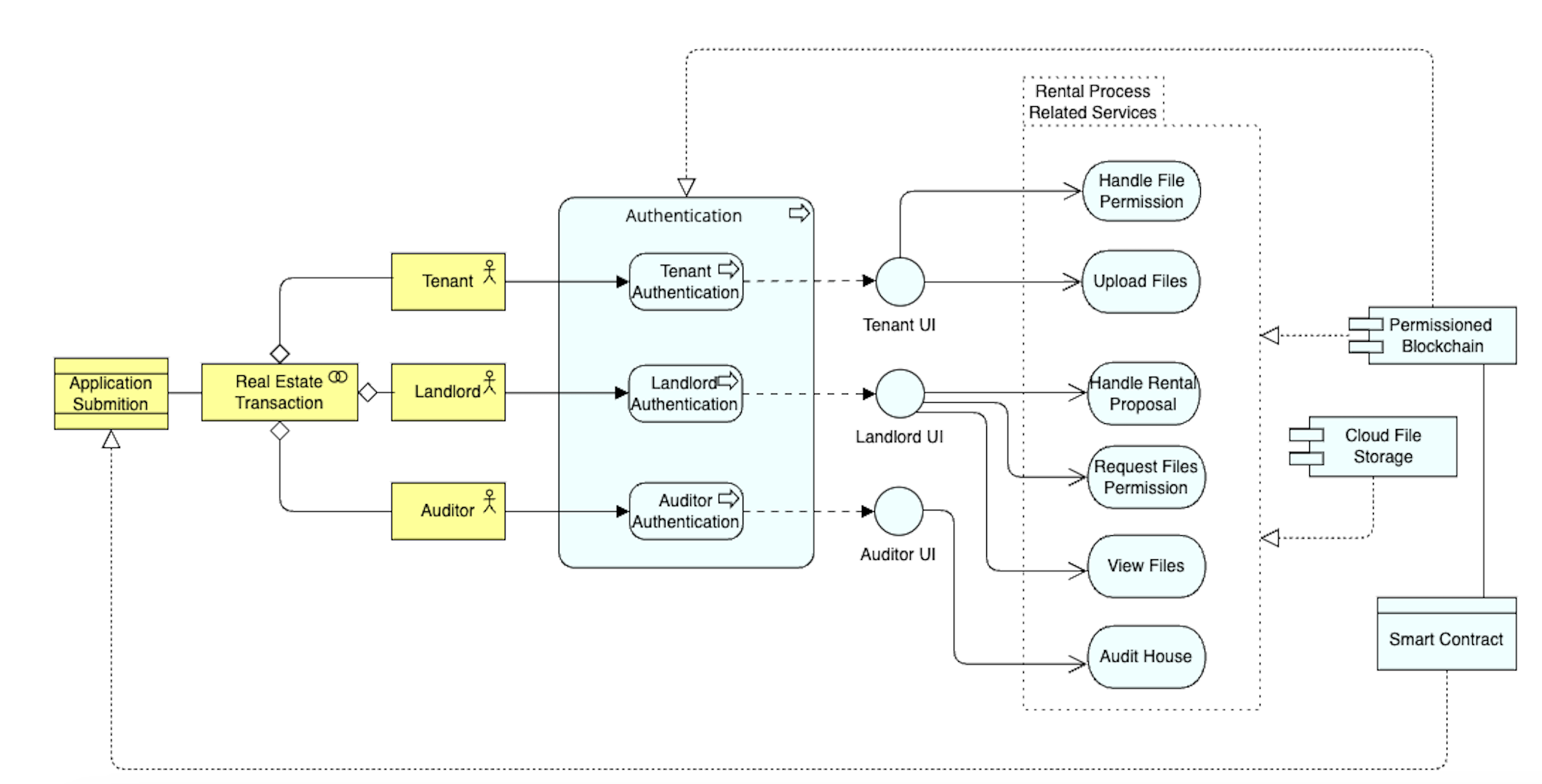}
\caption{Blockchain-Based Documentation Management Architecture Archimate diagram}
\label{architecture}
\end{figure}

The application architecture relies on multiple ecosystem components and a permissioned blockchain with a dedicated event register to support its core functionality. 
An illustration of this architecture can be observed in the figure provided \ref{architecture}.

\captionsetup[figure]
{font=footnotesize,labelfont=small,labelfont={bf}}

The given architecture comprises all the components required to make the BDM system work. The authentication module is a certificate authority that handles identities within a decentralised network. Every tenant that is authenticated on the platform must accept the terms and conditions so that, in the future, the system generates a final report with the data generated on the system. All related real estate rental services, document cloud storage, permissioned blockchain, and its smart contract are linked together to create an intelligent tool capable of auditing a house.

\begin{itemize}

    \item \textit{Authentication:} This component is the entry point in the application. It handles the identities of users using a certificate authority. 

    \item \textit{document Cloud Storage:} is the component where uploaded documents will be stored and accessed. 

    \item \textit{Permissioned Blockchain:} is a controlled and restricted blockchain network, limiting access to authorised participants. This solution will contain two organisations, the tenant organisation and the landlord/auditor organisation. Permissioned smart contracts save document-related information and handle authorisation to access documents stored in the document storage. The blockchain itself serves as the register, tracking every event, including permission grants and document uploads. 

    \item \textit{Smart Contract:} represents the core of this architecture, where specific techniques are applied to the document for each step of the process, defining a proper workflow for each task in the Rental Process Related Services.

    \item \textit{Rental Process Related Services:} are the methods established inside the smart contract that can be called depending on the role of the authenticated user.
    
\end{itemize}

An important concept for this architecture is the concept of a hash function \cite{preneel1994cryptographic}. It is a mathematical algorithm that takes an input and produces a fixed-length string of characters, known as the hash value or digest. It is designed to be a one-way process, which means that it should be computationally infeasible to reverse the hash value to retrieve the original input. The three main properties of a good hash function are as follows.

\begin{itemize}
    \item \textit{Deterministic:} the same input will always produce the same hash value.
    \item \textit{Fast computation:} it should be efficient to compute the hash for any given input.
    \item \textit{Collision Resistance} it is extremely unlikely for two different inputs to produce the same hash value.
\end{itemize}

The metadata of the uploaded documents will be stored in the smart contracts, and the document itself will be stored in the document cloud storage component. A Ricardian contract is a digital contract that combines a legal contract with a machine-readable contract, often used in blockchain technology to automate and verify contract terms \cite{grigg2004ricardian}. They allow the inclusion of legal language and privacy terms within the smart contract. This makes it possible to specify and enforce GDPR-related obligations, such as data protection, consent, and especially the right to be forgotten, directly in the contract code. To implement a Ricardian contract with a smart contract using a document hash:

\begin{enumerate}
    \item Create a legal document with terms and conditions.
    \item Compute the hash of the document.
    \item Embed the hash in the smart contract.
    \item Include logic to validate the document's hash in the smart contract.
    \item Implement digital signatures for consent.
    \item Deploy the smart contract on a blockchain.
    \item Maintain an audit trail of interactions on the blockchain.
\end{enumerate}

\subsection{System Processes}
\label{subsection: Application Sequence Diagrams}

The system processes outline the core processes within the application and highlight key interactions among tenants, landlords, and auditors. The three main processes covered include tenant document uploads, document authorisation, and the audit process. These processes are essential components of the application and encompass user onboarding, permission management, and regulatory compliance verification.

\subsubsection{Tenant Upload Documents Process}

The following Figure \ref{tenant_sequence} represents the first interaction with the application. A user with a tenant role registers and gets authenticated on the platform. Subsequently, he must accept the terms and conditions so that the system generates data for future audit requests. Then, he proceeds to choose a house to rent and submits its documents. The documents metadata are then uploaded to the blockchain, and the original documents are uploaded to the Google document storage. The next steps are discussed in Section \ref{Tenant and Landlord documents Permissions} where the tenant and the landlord interact for the handling of permissions on the uploaded documents.

\captionsetup[figure]
{font=footnotesize,labelfont=small,labelfont={bf}}
\begin{figure}[h!]
\centering
\includegraphics[width=0.45\textwidth]{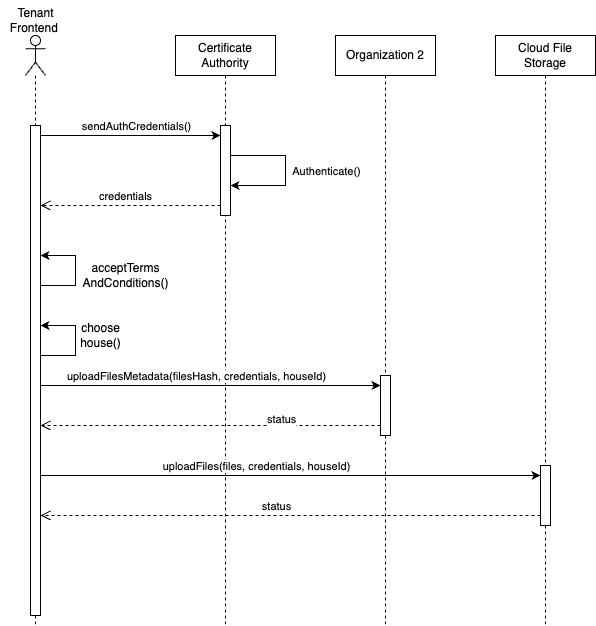}
\caption{Tenant documents Upload Process Sequence Diagram}
\label{tenant_sequence}
\end{figure}

\subsubsection{Document Authorisation Process}
\label{Tenant and Landlord documents Permissions}

\begin{figure}[h!]
\centering
\includegraphics[width=0.5\textwidth]{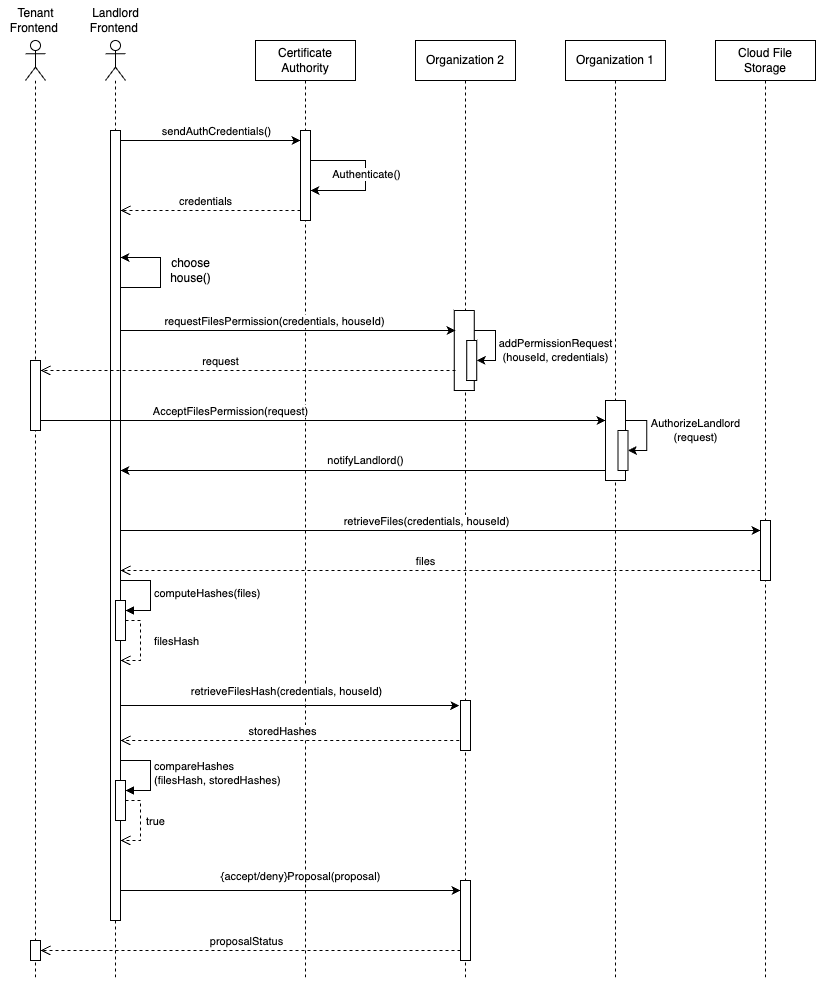}
\caption{document Authorisation Process Sequence Diagram}
\label{landlord_sequence}
\end{figure}

Figure \ref{landlord_sequence} illustrates the interaction following the submission of a tenant's proposal. The landlord, authenticated on the platform, selects a house with pending applications. He requests permission to view the original proposal documents, initiating a notification to the tenant. The tenant can either accept or decline the request. Upon acceptance, the landlord retrieves the original documents from Google Cloud storage. After reviewing documents from multiple applications, the landlord selects the best proposal, updating the proposal status for the tenant organization.

For document integrity verification, the system retrieves the document's hash from the blockchain and compares it to the hash generated from the document in storage. A match confirms document integrity, while a mismatch indicates corruption. The landlord can then proceed to accept or deny the proposal application.

\captionsetup[figure]
{font=footnotesize,labelfont=small,labelfont={bf}}

\subsubsection{Auditing Process}
\label{auditor_sequence_subsubsection}

\begin{figure}[h]
\centering
\includegraphics[width=0.5\textwidth]{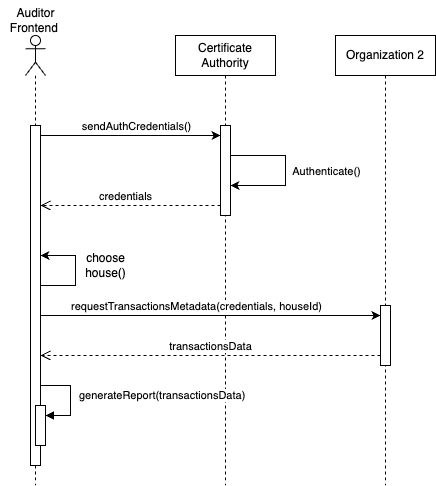}
\caption{Auditing Process Sequence Diagram}
\label{auditor_sequence2}
\end{figure}

The final possible interaction of the application occurs after the sequence \ref{landlord_sequence}. Now that the system has transactions, a verified auditor can request to audit a chosen house and verify that everything is according to the regulations. The sequence \ref{auditor_sequence2} starts by authenticating the auditor. An house is chosen, and a request to retrieve the documents metadata is made. Then the frontend generates a final report for the auditor to download.

\captionsetup[figure]
{font=footnotesize,labelfont=small,labelfont={bf}}

\section{Blockchain-Based Documentation Management Implementation}
\label{chapter:Blockchain Based Document Management}

In implementing this specific use case, an approach has been developed that leverages the features of the Hyperledger Fabric Framework to seamlessly integrate blockchain technology. This implementation is further supported by the creation of a frontend application using React, a JavaScript framework. This introduction lays the foundation for a thorough examination of the implemented approach and its essential components.

\subsection{Blockchain Architecture}
\label{subsubsection: blockchain architecture}

\begin{figure}[h]
\centering
\includegraphics[width=0.5\textwidth]{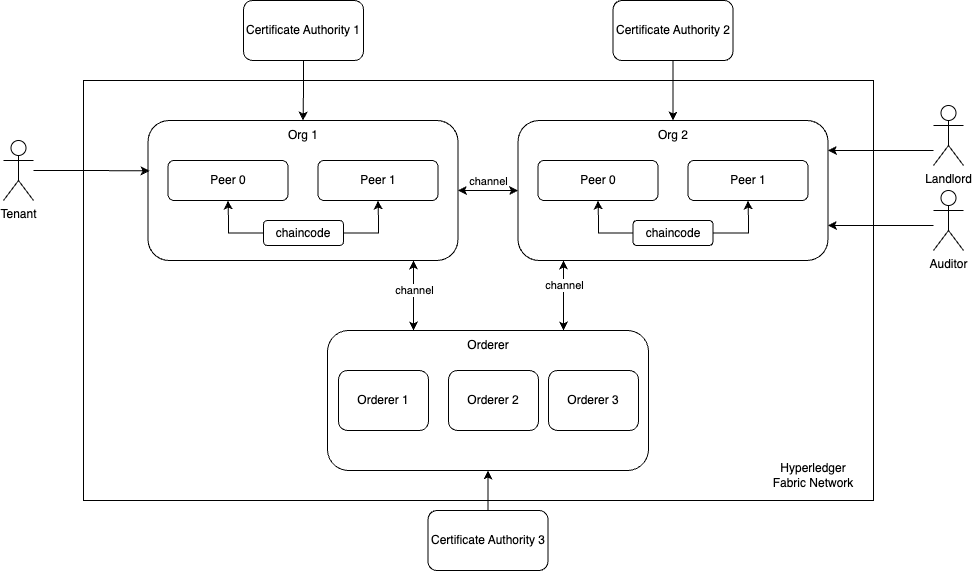}
\caption{Hyperledger Fabric Implemented Network Architecture}
\label{hlf_architecture}
\end{figure}

The network comprises distinct components: certificate authorities, organisations, an orderer, and channels, each with a defined role in the overall structure. These elements are explained in the following bullet points, and the blockchain infrastructure can be seen in Figure \ref{hlf_architecture}

\captionsetup[figure]
{font=footnotesize,labelfont=small,labelfont={bf}}

\begin{itemize}
    \item \textit{Organization:}  Organisations in this context are entities that define the participants in the network. Each organisation typically has its own set of peer nodes, a Certificate Authority (CA), and administrative control over its members.

    \item \textit{Certificate Authority:} The CA issues digital certificates to network participants. These certificates contain cryptographic keys and are used to verify the identity of the nodes and maintain the integrity and confidentiality of transactions \ref{section:Authentication}.

    \item \textit{Chaincode:} Chaincode or Smart Contract, is a piece of code that defines the rules and logic for transactions on the blockchain. It is installed on peer nodes and can be invoked to modify the ledger state. Before chaincode can be used, it must go through an approval and commitment process. This involves an endorsement policy, where peers validate and approve the chaincode, and a commitment to the channel ledger. This ensures that all organisations agree on the legitimacy of the code.

    \item \textit{Peer Nodes:} Peer nodes are individual instances within an organisation that maintain a copy of the ledger. They execute chaincode transactions, validate transactions, and endorse them before they are added to the blockchain. Having two peer nodes per organisation ensures redundancy and high availability.

    \item \textit{Orderer Nodes:} Orderer nodes are responsible for maintaining the order of transactions on the blockchain. They validate transactions, create blocks, and ensure consensus among network participants. The consensus algorithm can be crash-fault-tolerant or byzantine-fault-tolerant. Three orderer nodes enhance fault tolerance and maintain the integrity of the ledger.

    \item \textit{Channel:} A channel is a private communication layer in the blockchain network that allows the segregation of transaction data. It restricts access to specific organisations, ensuring that only authorised participants can view and transact on this channel.
    
    \item \textit{Network APIs:} Application Programming Interfaces provide an interface for external applications to interact with the blockchain network. In this case, two APIs are deployed, each tailored to a specific organisation, allowing authorised users to send transactions and retrieve data from the blockchain.
    
\end{itemize}

In this architecture, tenants connect to Organisation 1 to input data into the ecosystem, while landlords and auditors connect to Organisation 2. They use this connection to view uploaded files or generate reports based on information extracted from blockchain transactions. Every transaction is signed by its author, and therefore non-repudiation is granted.

\subsection{Authentication}
\label{section:Authentication}

Within the context of user identity management in Hyperledger Fabric, there exists a structured process to enable secure participation in the network. This process involves user registration, during which individuals provide vital information, such as their username, password, and role. After successful registration, users are equipped with cryptographic credentials, namely an X.509 certificate and a private key. These credentials establish their secure digital identity within the network. The user identity is securely stored in a wallet, protecting cryptographic keys and certificates from unauthorised access. When users intend to log in, the verification of their identity takes place through the Certificate Authority (CA), using the certificate and private key stored in the wallet for authentication. This meticulous process ensures that only authorised users, possessing valid credentials, gain access to the blockchain network, thus ensuring the security and reliability of interactions.

\begin{itemize}
    \item \textit{Registration with the Certificate Authority:} The code includes a registration process that allows new users to join the Hyperledger Fabric network securely. When a user wishes to register, they provide essential information, such as a username, password, and role. The code first checks if the user identity already exists within the CA. If not found, it proceeds with the registration. During registration, the user enrolment ID and secret, often chosen by the user during signup, are used. These credentials are crucial to authenticating the user within the network.

    \item \textit{Enrolment and Identity Creation:} Following successful registration, the code initiates the enrolment process. This step involves obtaining cryptographic credentials for the user, namely, an X.509 certificate and a private key. These credentials serve as the user digital identity within the Hyperledger Fabric network. The enrolment process ensures that the user identity is securely generated and linked to the CA. This identity creation process is an integral part of ensuring secure and authenticated interactions with the blockchain network.

    \item \textit{Storage in the Wallet:} Once the user identity is generated and enroled with the CA, it is securely stored in a wallet. The wallet acts as a secure repository for user identities. It ensures that cryptographic keys and certificates are protected from unauthorised access. Users can conveniently access their identities from the wallet for subsequent interactions with the network. This secure storage mechanism is vital to maintaining the confidentiality and integrity of user credentials.

    \item  \textit{Log in with the CA and Wallet:} When a user wants to log in, the code checks the CA to verify the user identity. If the identity is found, the user X.509 certificate and private key stored in the wallet are used for authentication. This login process ensures that only authorised users with valid credentials can access the blockchain network. It also provides a secure and convenient way for users to participate in blockchain transactions and queries.
\end{itemize}

The authentication phase provides security by generating and storing user identities, ensuring that only authorised users can interact with the Hyperledger Fabric network. These processes are essential to maintain the integrity and confidentiality of blockchain transactions and user data.

\subsection{Cloud File Storage}
\label{section:Cloud Storage}

Google Cloud File Storage was selected for file storage and retrieval based on personal experience with the technology. A Google Service Account is essential for secure and automated access to Google Cloud services, enabling the application to interact with Google resources without user passwords. This is crucial for data processing, server-to-server communication, and integrating the application frontend with Google Cloud services. The main storage features include:

\subsubsection{Store Files}

\begin{itemize}
    \item \textit{Upload Files:} upload files to Google Cloud File Storage using Google Service Account credentials through the Web interface.

    \item \textit{Organize Files:} organise files into folders, all managed by the Google Service Account, ensuring a well-maintained and structured storage system.

    \item \textit{Permission settings:} precise control over access permissions, granting read-only, read-write, or customised access to specific users or groups through the Google Service Account.

\end{itemize}

\subsubsection{Retrieve Files}

\begin{itemize}
    \item \textit{Access Anywhere:} retrieve stored files from Google Cloud Storage using the Google Service Account within the Web interface.

    \item \textit{Search and Retrieve:}  locate files within the React application by employing keywords or parameters in the search function.

    \item \textit{Permission settings:} enforce strict access control during file retrieval, allowing only authorised users authenticated by the Google Service Account to view or modify files.

\end{itemize}

\subsection{Smart Contract Implementation}
\label{subsection: smart_contract_solution}

The application's core functions are divided into write and read functions, enabling users to interact with documents, houses, proposals, and access requests. These functions are crucial to creating, managing, and retrieving historical data, ensuring effective and secure user interactions. 

\subsubsection{Implemented Functions}

\begin{itemize}
    \item \textit{createHouse:} Allows users to create a new house associated with a landlord, subject to necessary permissions and checks.

    \item \textit{createProposal:} Enables tenants to create rental proposals for houses and landlords, following permission and existence checks.
    
    \item \textit{denyProposal:} Permits landlords to reject tenant rental proposals for specific houses, after verifying permissions and existence.
    
    \item \textit{acceptProposal:} Allows landlords to accept tenant rental proposals, subject to checks and permissions.
    
    \item \textit{getRequestsForTenant:} Retrieves access requests made by tenants after ensuring caller existence and permissions.
    
    \item \textit{createDocument:} Tenants create documents related to rented houses, provided they have the necessary permissions and meet house-related criteria.
    
    \item \textit{requestAccess:} Tenants request access to specific documents from landlords, subject to various checks.
    
    \item \textit{acceptAccess:} Tenants grant access to landlords for specific documents, ensuring permissions and existence.
    
    \item \textit{denyAccess:} Tenants deny access to landlords for specific documents, following checks and permissions.
    
    \item \textit{getDocument:} Allows users to retrieve document details with proper access rights and after confirming document existence and permissions.

    \item \textit{getProposalsForLandlord:} Retrieves rental proposals made to a specific landlord, provided that the landlord exists and has the necessary permissions.
    
    \item \textit{getHistoricData:} Offers comprehensive historical data, including document metadata history, proposal history, and access request history. It checks various conditions for each aspect to ensure that data retrieval is valid and secure.
\end{itemize}

\subsection{User Interfaces}

For the different types of individual who interact with the system, a frontend has been developed with three distinct interfaces: one for tenants, one for landlords, and one for auditors. The frontend is developed in React version v18.2.0, a JavaScript framework chosen for its flexibility and performance.

These interfaces provide a simple way for tenants to apply for properties and manage them, landlords to manage their properties and tenant applications, and auditors to review every transaction for a certain house rental.

\captionsetup[figure]
{font=footnotesize,labelfont=small,labelfont={bf}}
\begin{figure}[h]
\centering
\includegraphics[width=0.5\textwidth]{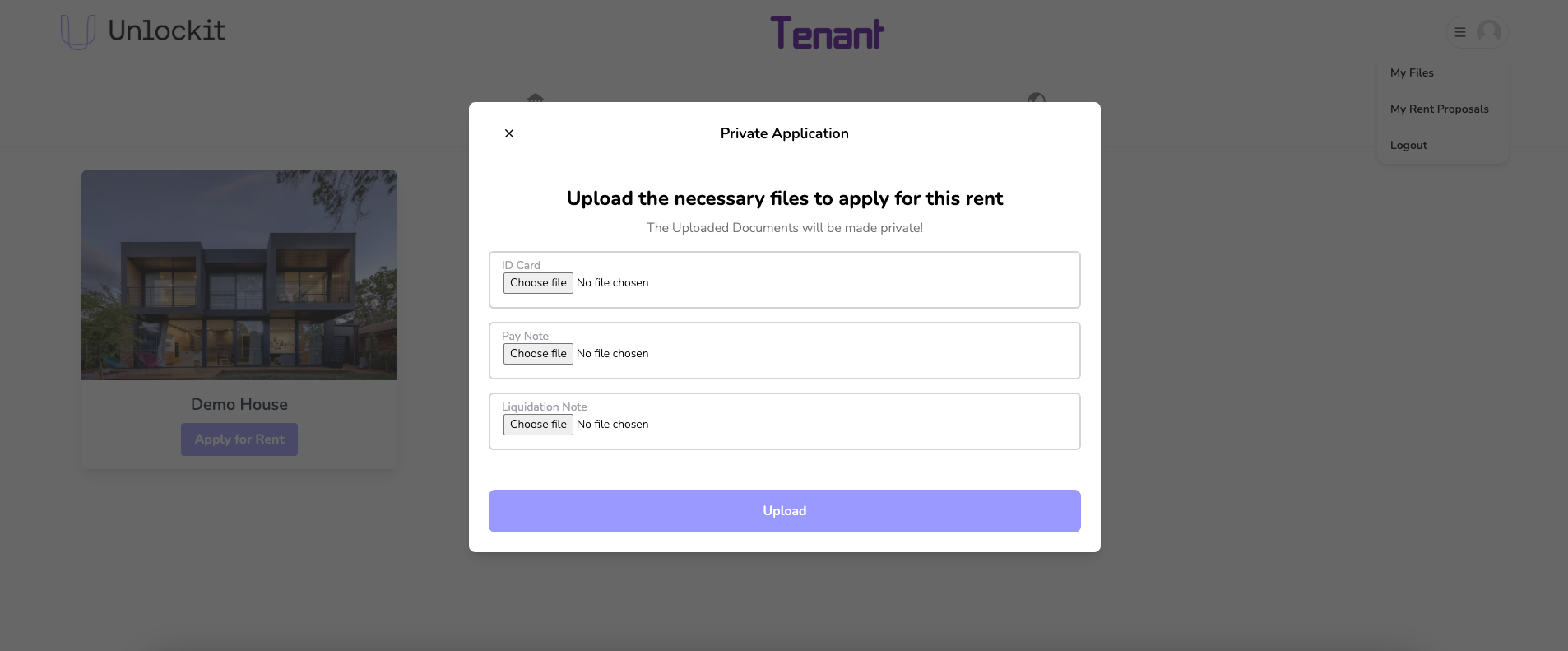}
\caption{Tenant Document Upload View}
\label{Tenant Document Upload View}
\end{figure}

Figure \ref{Tenant Document Upload View} displays the Tenant Document Upload View, where tenants can securely upload the necessary documents for their rental application. This figure illustrates the document management capabilities of the interface, which streamlines the process for tenants while ensuring the secure storage of important documentation.

\captionsetup[figure]
{font=footnotesize,labelfont=small,labelfont={bf}}
\begin{figure}[h]
\centering
\includegraphics[width=0.5\textwidth]{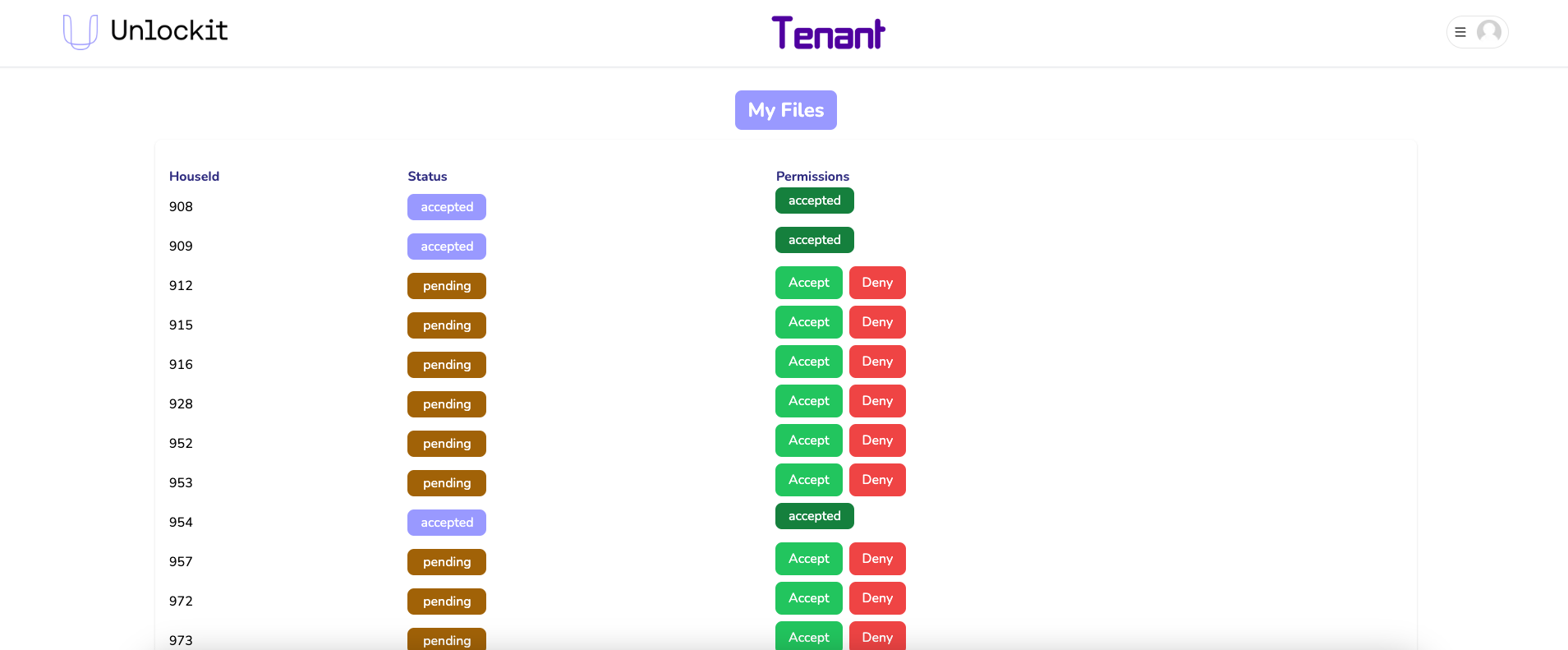}
\caption{Tenant Document Authorisation View}
\label{Tenant Document Authorisation View}
\end{figure}

The Tenant Documents Authorisation View, as shown in Figure \ref{Tenant Document Authorisation View}, where possible tenants manage and authorise access to their uploaded documents. This figure hints at features that allow tenants to grant access to these documents to landlords or other authorised parties, maintaining control over their information.

\captionsetup[figure]
{font=footnotesize,labelfont=small,labelfont={bf}}
\begin{figure}[h]
\centering
\includegraphics[width=0.5\textwidth]{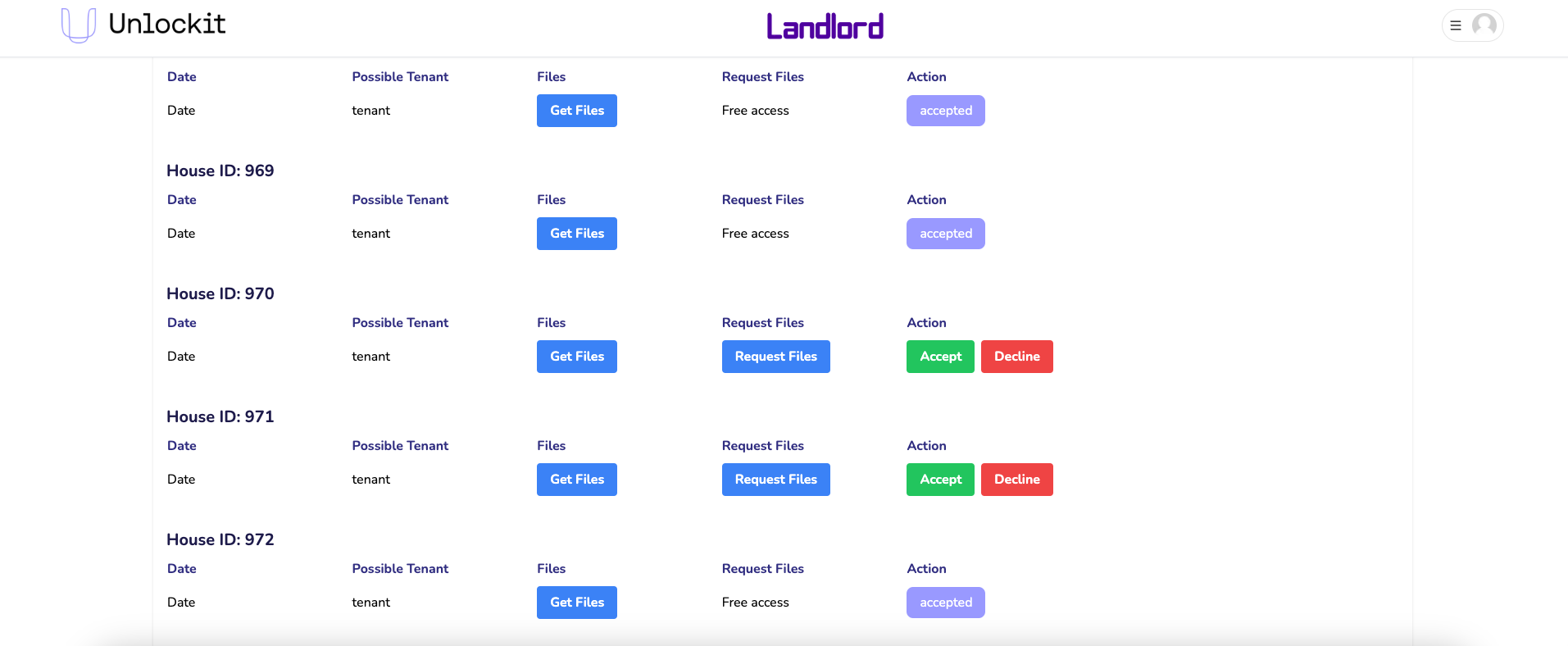}
\caption{Landlord Renal Proposals View}
\label{Landlord Renal Proposals View}
\end{figure}

Figure \ref{Landlord Renal Proposals View} presents the Landlord Rental Proposals View, which showcases the interface designed for landlords. This interface likely enables landlords to review tenant applications and manage rental proposals efficiently. Provides a comprehensive overview of tenant applications, helping landlords in their decision-making process.

\subsection{Audit Report}

The blockchain system's audit report meticulously records all transactions related to a specific house, offering a unique capability to track the rental process. This transparency and precision simplify future audits. The built-in Hyperledger Fabric function, \textit{getHistoryByKey} is used to achieve this functionality. It provides the complete history of an object using a key representing a house, its landlord, and tenants. Accessing the history of the \textit{house} object reveals the entire transaction lifecycle of that specific rental.

\begin{figure}[h]
\centering
\includegraphics[width=0.5\textwidth]{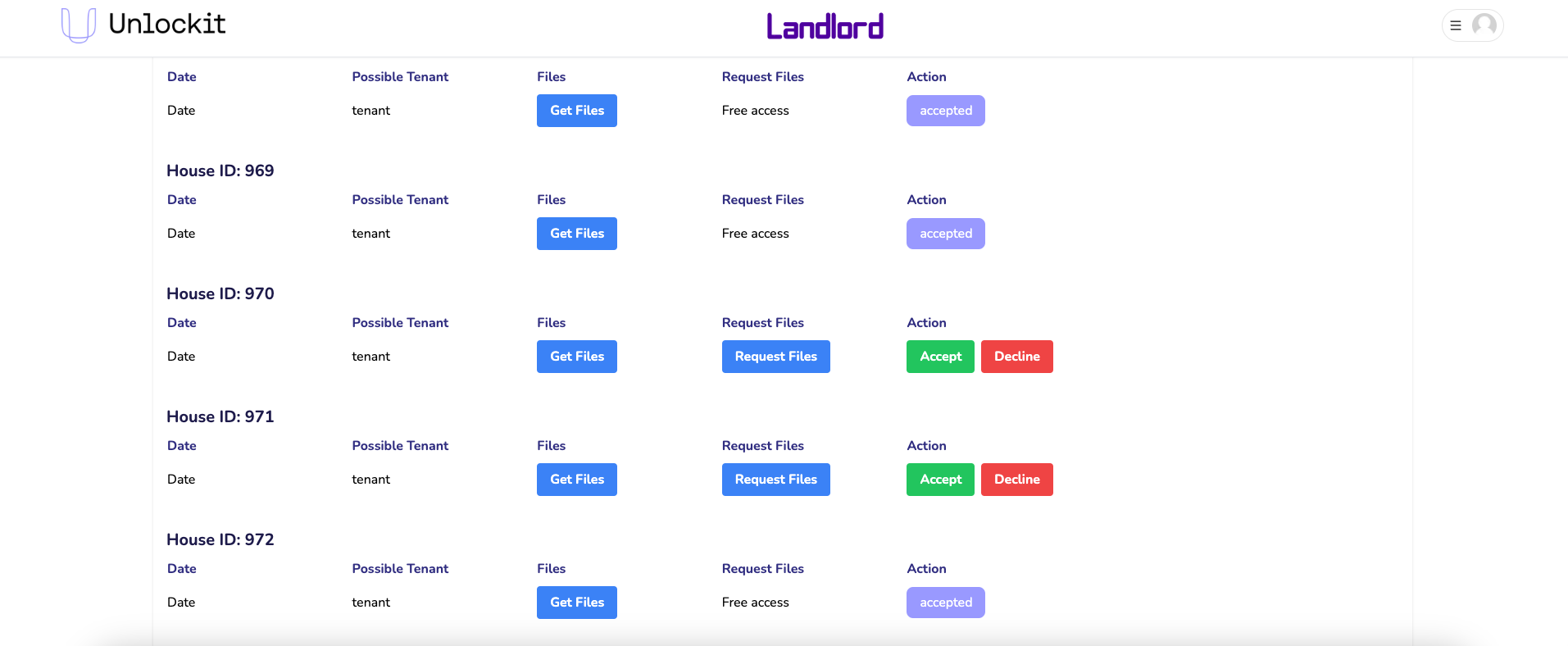}
\caption{Landlord Renal Proposals View}
\label{fig:audit_report}
\end{figure}

Authorised auditors can obtain detailed records of the rental process, from application to approval. Figures \ref{fig:audit_report} presents a simple audit report for a rental application.

\captionsetup[figure]
{font=footnotesize,labelfont=small,labelfont={bf}}

\section{Evaluation}
\label{chapter:results}

\subsection{Methodology}

Apache JMeter is a versatile tool for testing server-based applications. It provides features like result trees and aggregate reports to analyse performance. The result trees show the details of the execution of the HTTP request, which aids in the identification of the issue. Aggregate reports compile metrics across multiple test runs, revealing insights into response times, throughput, and errors.

\subsection{Experimental Setup}

\captionsetup[figure]
{font=footnotesize,labelfont=small,labelfont={bf}}
\begin{figure}[h]
\centering
\includegraphics[width=0.5\textwidth]{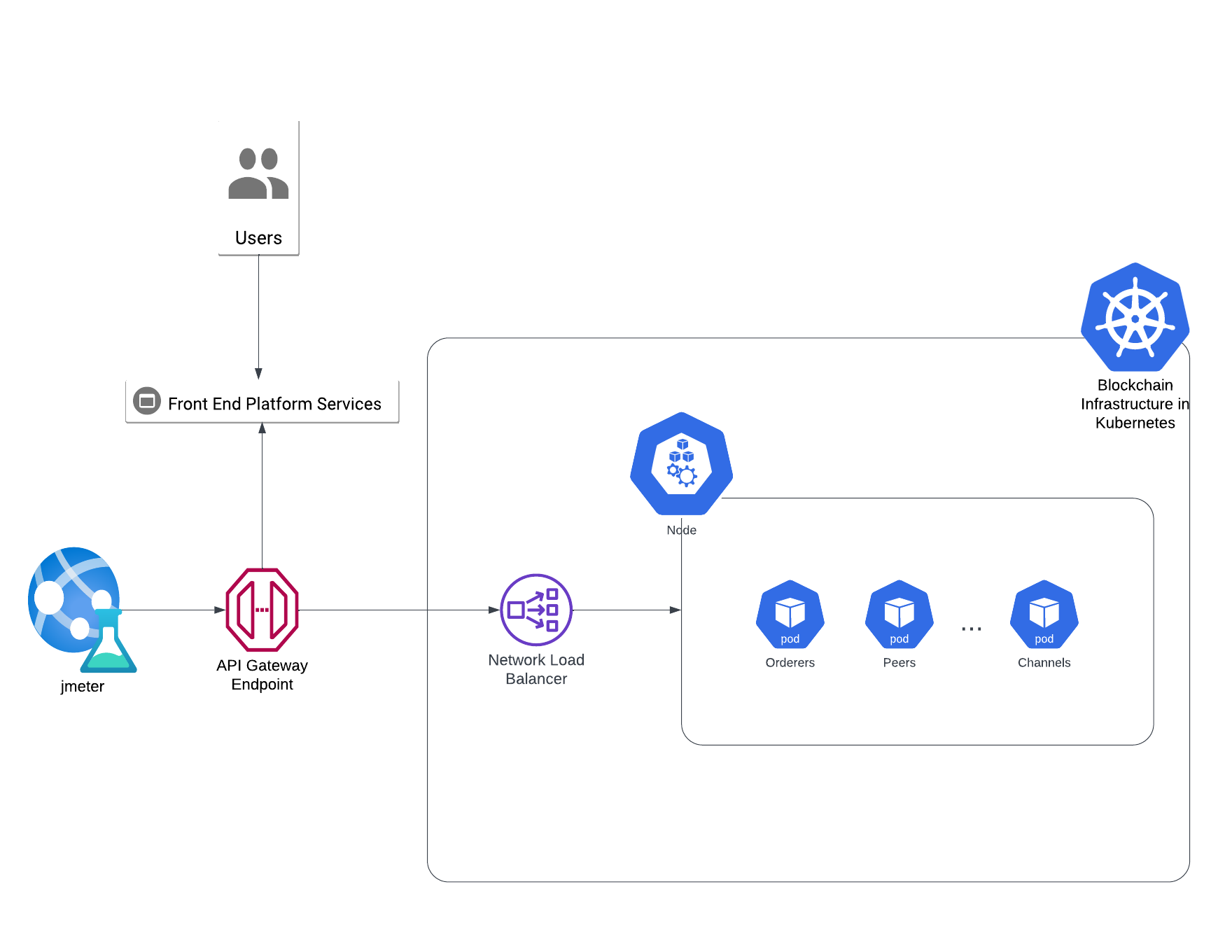}
\caption{Blockchain-based Documentation Management Kubernetes Diagram}
\label{Kubernetes}
\end{figure}

Figure \ref{Kubernetes} presents the Hyperledger Fabric network infrastructure, hosted within a locally deployed Kubernetes cluster. This deployment relies on KinD (Kubernetes in Docker), a tool that facilitates the creation and management of Kubernetes clusters using Docker containers as nodes. The process is streamlined through the use of the HLF Operator, a Kubernetes operator designed to simplify the deployment and management of Hyperledger Fabric networks within Kubernetes clusters.

\subsection{Experimental Results}

Evaluation of system performance provided critical insight into how the system behaves under varying loads of concurrent requests. Eight functions were examined, with request loads ranging from 50 to 1000 concurrent requests. A consistent trend emerged: As the number of concurrent requests increased, the system latency increased, leading to decreased throughput. This well-documented inverse relationship between latency and throughput was observed. The following Figures \ref{fig:latency} and \ref{fig:throughput} present specific findings related to individual functions, revealing their strengths and vulnerabilities with increasing concurrent requests. These results serve as a basis for optimising the system for real-world scenarios with dynamic workloads.

\captionsetup[figure]
{font=footnotesize,labelfont=small,labelfont={bf}}
\begin{figure}[h]
\centering
\includegraphics[width=0.5\textwidth]{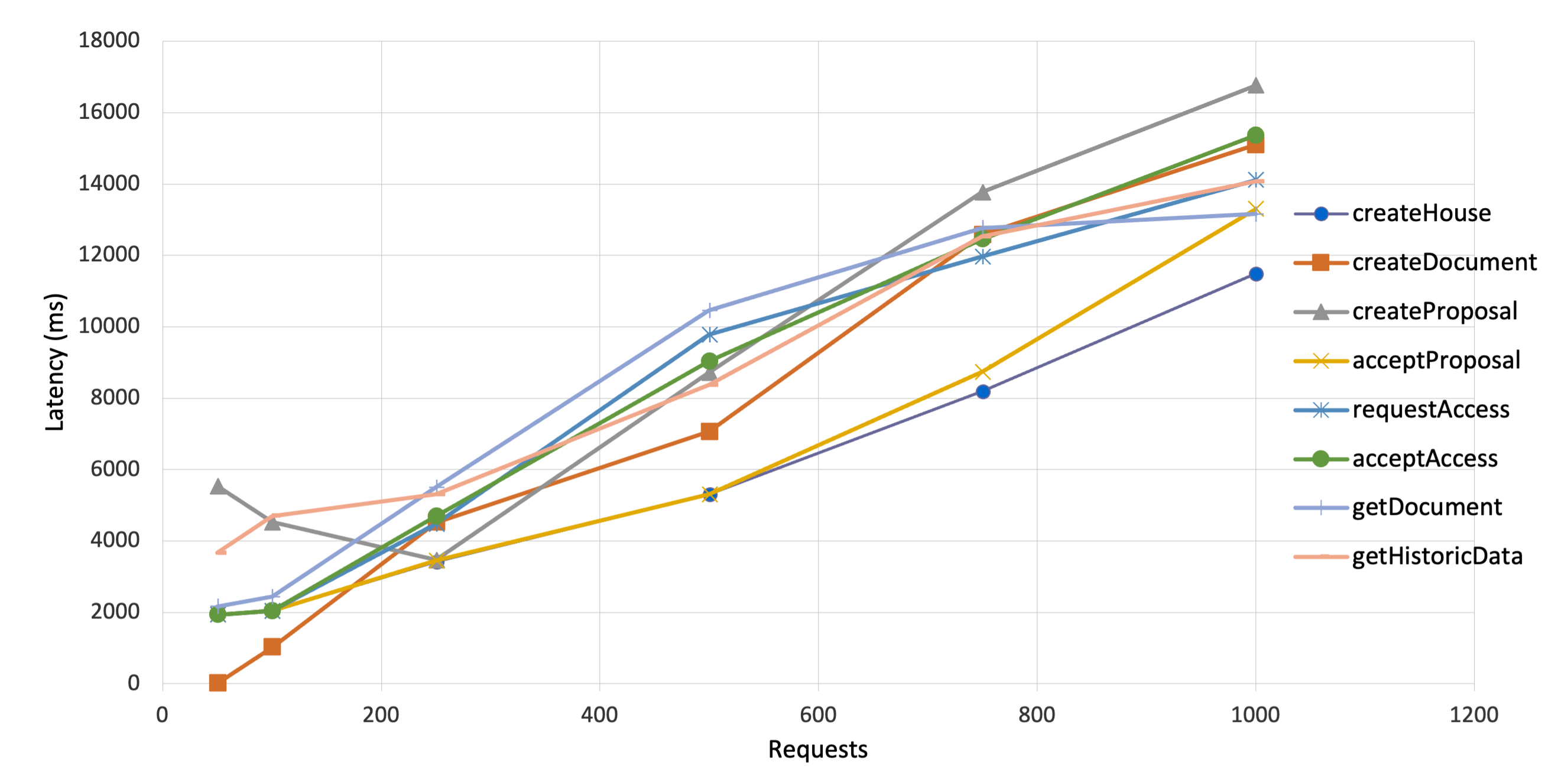}
\caption{Latency for each Implemented Function}
\label{fig:latency}    
\end{figure}

\captionsetup[figure]
{font=footnotesize,labelfont=small,labelfont={bf}}
\begin{figure}[h]
\centering
\includegraphics[width=0.45\textwidth]{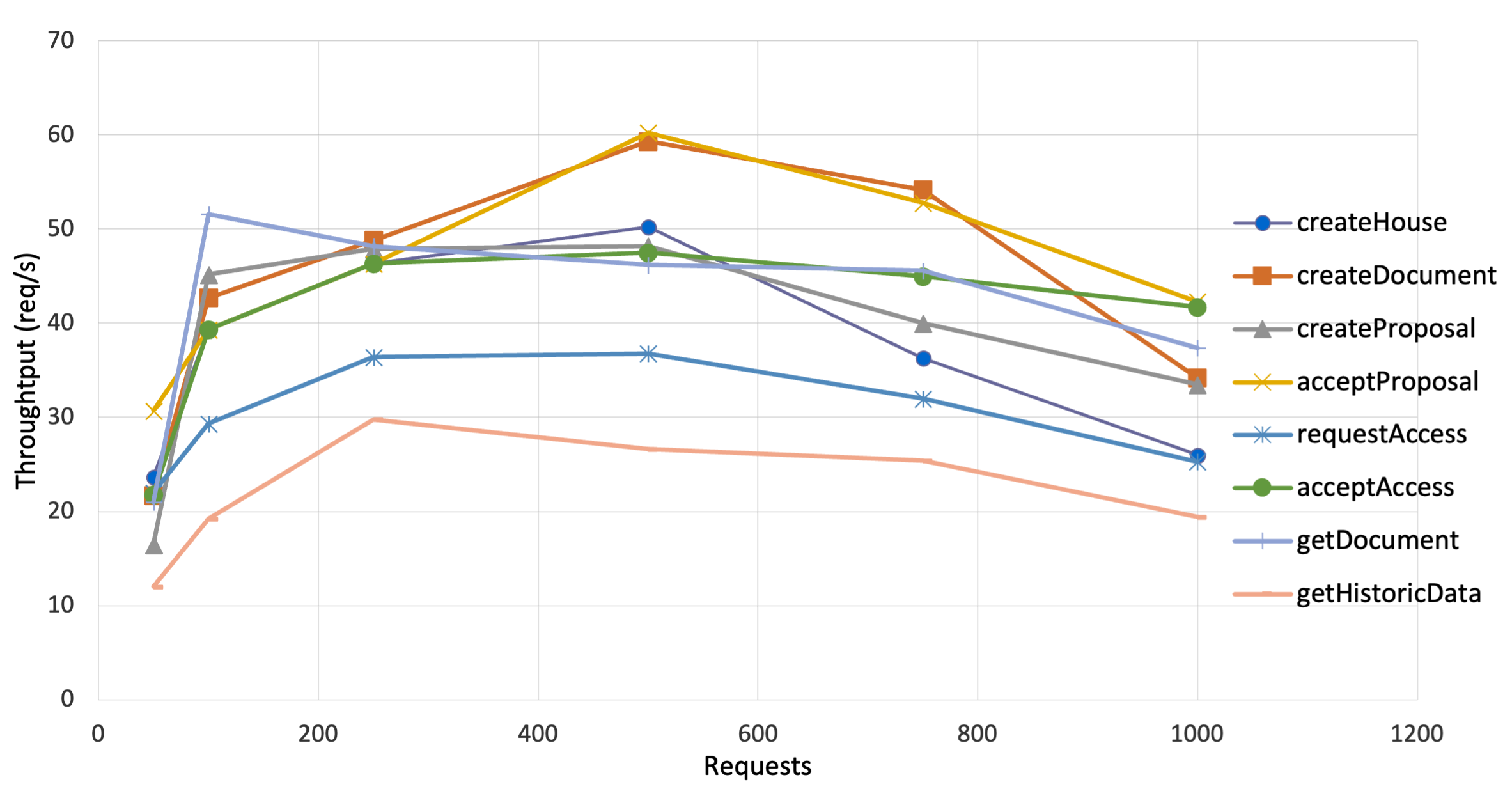}
\caption{Throughput for each Implemented Function}
\label{fig:throughput}    
\end{figure}

As expected, the results demonstrate that as concurrent requests increase, system latency increases while throughput decreases. The increase in latency with higher concurrency can be attributed to the system finite resources being spread among multiple requests, resulting in delays in processing, leading to higher response times. Concurrently, reduced throughput indicates that the system handles fewer requests per second as concurrency levels increase. This is due to the longer time required to process each request, which limits the overall capacity of the system.
At the latency level, the functions constitute more or less the same latencies per number of requests, with the createProposal function being the slowest one. The getHistoricData function is the one with the lowest throughput, as expected, since it must carry for each historic transaction its content, resulting in the usage of more bandwidth per request. On the other hand, acceptAccess and acceptProposal are the ones with the highest throughput for 1000 requests, since they require less data. At high request levels, the system encounters issues like denying or dropping connections, impacting the above results, leading to request exceptions. These exceptions cause some requests to terminate prematurely, resulting in response times shorter than anticipated. For a comprehensive overview of system malfunction across functions and concurrency levels, Table \ref{error table} is provided. 

\begin{table}[h]
\huge
\centering
\begin{adjustbox}{width=0.5\textwidth}
\begin{tabular}{c|c|c|c|c|c|c|c|c|}
\cline{2-9}
                           & createHouse & createDocument & createProposal & acceptProposal & requestAccess & acceptAccess & getDocument & getHistoricData \\ \hline
\multicolumn{1}{|c|}{50}   & 0           & 0              & 0              & 0              & 0             & 0            & 0           & 0               \\ \hline
\multicolumn{1}{|c|}{100}  & 0           & 0              & 0              & 0              & 0             & 0            & 0           & 0               \\ \hline
\multicolumn{1}{|c|}{250}  & 0           & 0              & 0              & 0              & 0             & 0            & 0           & 0               \\ \hline
\multicolumn{1}{|c|}{500}  & 0           & 0              & 0              & 0              & 0             & 0            & 0           & 0               \\ \hline
\multicolumn{1}{|c|}{750}  & 0.15        & 78.53          & 0              & 0              & 0.6           & 0.6          & 0           & 2.3             \\ \hline
\multicolumn{1}{|c|}{1000} & 2           & 80             & 0              & 0              & 5             & 4            & 2           & 12              \\ \hline
\end{tabular}
\end{adjustbox}
\caption{Error Percentage considering the number of requests for each function}
\label{error table}
\end{table}

Most functions begin to encounter errors after around 750 requests. The \textit{createDocument} function has a notably higher error rate compared to other functions because it handles the storage of document metadata, resulting in larger data transactions per request. In contrast, \textit{createProposal} and \textit{acceptProposal} functions have no errors, as they are simpler and involve smaller data transactions. \textit{createHouse}, \textit{requestAccess}, \textit{acceptAccess}, and \textit{getDocument} functions exhibit low error levels even at 1000 requests. However, \textit{getHistoricData} has a 12\% error rate of 1000 requests due to larger data responses, occasionally causing connection drops.

\subsection{Current Process vs Smart Audit Process}

The Blockchain-based documentation management with audit support system represents an advance over the current manual processes of IMPIC. In the current system, auditors spend a substantial amount of time and effort manually collecting, validating, and reconciling data from various sources. This is a laborious and time-consuming task. However, with the introduction of blockchain, these tasks are performed on top of verified data, freeing auditors from repetitive work and enabling them to focus on more strategic aspects of auditing.

Furthermore, the use of blockchain technology ensures the accuracy and reliability of the data. In the manual system, the risk of human errors is a constant concern that requires additional effort to verify the data. On the contrary, the blockchain immutable ledger guarantees the integrity of information, reducing the need for extensive error checking and increasing efficiency.

Collaboration and transparency are greatly improved by the blockchain-based platform. Current collaboration methods with stakeholders can be malicious and lack transparency. However, the system offers a secure platform for all parties to contribute and access data transparently with the necessary authorisation, facilitating seamless communication and improving transparency.

In addition, the system supports security and compliance efforts. Ensuring data security and compliance can be challenging in the current system. However, the cryptographic security features of the blockchain and audit trail capabilities improve data security and compliance, reducing potential legal and financial risks.

Finally, the system results in significant time and cost savings. Manual processes are resource-intensive and can lead to high costs. Automation, real-time data access, and improved accuracy result in significant time and cost savings for auditors, allowing them to allocate resources more efficiently and strategically.

\section{Conclusion}

This document introduces a solution to streamline the house rental process and audit reporting in the real estate market, with the aim of enhancing transparency and efficiency for tenants, landlords, and auditors. It leverages blockchain technology for added security and efficiency. In the real estate rental market, three main parties are involved: tenants, landlords, and auditors. Each party has specific functions within the system. Blockchain technology is explored as a means to improve the real estate market by offering decentralisation. The document introduces a blockchain-based documentation management system that improves efficiency and security. The solution proposes the use of a permissioned blockchain to provide controlled access to audit data for tenants, landlords, and auditors while ensuring strict access controls. Key functions in the system include creating and accepting rental proposals, accessing and verifying documents, and maintaining the integrity of uploaded files through hashing. Performance evaluation of the system involved measuring latency, throughput, and errors. The results indicate the system's capability to handle up to 500 concurrent transactions without errors. Additionally, the document discusses how this technology can make auditors more efficient by automating repetitive tasks, comparing it with the existing processes used by IMPIC.

\section{Acknowledgements}

This work was developed within the scope of the project nr.\ 51 ``BLOCKCHAIN.PT - Agenda Descentralizar Portugal com Blockchain'', financed by European Funds, namely ``Recovery and Resilience Plan - Component 5: Agendas Mobilizadoras para a Inova\c{c}\~{a}o Empresarial'', included in the NextGenerationEU funding program.  This work was also supported by national funds through Funda\c{c}\~{a}o para a Ci\^{e}ncia e a Tecnologia (FCT) with reference UIDB/50021/2020 (INESC-ID).



\end{document}